\documentclass[twocolumn,showpacs,preprintnumbers,amsmath,amssymb]{revtex4}

\usepackage{graphicx}
\usepackage{dcolumn}
\usepackage{bm}
\def\jpsi{{J/\psi}}

\def\be{\begin{equation}}
\def\ee{\end{equation}}
\def\bea{\begin{eqnarray}}
\def\eea{\end{eqnarray}}
\def\NO{\nonumber}
\def\gev{\mathrm{~GeV}}

\def\dfrac{\displaystyle\frac}

\def\md{\mathrm{d}}

\def\a{\alpha}
\def\b{\beta}

\def\d{\delta}
\def\e{\epsilon}
\def\g{\gamma}
\def\s{\sigma}

\begin{document}


\title{QCD corrections to $\jpsi$ polarization of hadronproduction at Tevatron and LHC}

\author{Bin Gong and Jian-Xiong Wang}%
\affiliation{
Institute of High Energy Physics, Chinese Academy of Sciences, P.O. Box 918(4), 
Beijing, 100049, China.\\
Theoretical Physics Center for Science Facilities, Beijing, 100049, China.
}%
\date{\today}

\begin{abstract}
The next to leading order (NLO) QCD corrections to $\jpsi$ polarization
of hadronproduction at Tevatron and LHC are calculated. The results show that the $\jpsi$ polarization is extremely changed from more transversal polarization at leading 
order (LO) into more longitudinal polarization at NLO. Although it gives more 
longitudinal polarization than the recent experimental result on the $\jpsi$ 
polarization at Tevatron. It sheds light on the solution to the large discrepancy 
of $\jpsi$ polarization between theoretical predication and experimental measurement,
and suggests that the next important step is to calculate the NLO correction for 
color octet state $\jpsi^{(8)}$ hadronproduction. Our calculations are performed 
in two ways where the polarizations are summed analytically or not, and they are checked with each other. It also gives a K factor for total cross section (ratio of NLO to LO) of about 2 and shows that the NLO corrections boost the $\jpsi$ production for about 2 order of magnitude in high transverse momentum $p_t$ region of $\jpsi$, 
which confirms the calculation by Campbell, Maltoni and Tramontano. 
\end{abstract}

\maketitle

The study of $J/\psi$ production on various experiments is a very interesting topic since its discovery in 1974. It is a good place to probe both perturbative and 
nonperturbative aspects of QCD dynamics. To describe the huge discrepancy of 
the high-$p_t$ $J/\psi$ production between the theoretical calculation and 
the experimental measurement, color-octet mechanism\cite{Braaten:1994vv} was proposed based on the non-relativistic QCD(NRQCD)\cite{Bodwin:1994jh}. The factorization formalism of NRQCD provides a theoretical framework to the treatment of heavy-quarkonium production. It allows consistent theoretical prediction to be made and to be improved perturbatively in the QCD coupling constant $\a_s$ and the heavy-quark relative velocity $v$.
Although it seems to show qualitative agreements with experimental data, there
are certain difficulties in the quantitative estimate in NRQCD for $J/\psi$ and 
$\psi'$ photoproduction at the DESY ep collider HERA
\cite{Ko:1996xw,kramer:1995nb},
$J/\psi(\psi')$ polarization of hadronproduction at
the Fermilab Tevatron,
and $\jpsi$ production in B-factories.

There are a few examples shown that NLO corrections are quite large 
and it is difficult to obtain agreement between the 
experimental results and leading order theoretical predictions for $J/\psi$ production.   
It was found that the current experimental results on 
inelastic $J/\psi$ photoproduction are adequately 
described by the color singlet channel alone once higher-order QCD corrections 
are included\cite{kramer:1995nb}.
Ref.~\cite{Klasen:2001cu} found that the DELPHI
\cite{deBoer:2003xm} data evidently favor the NRQCD formalism for
$J/\psi$ production $\gamma  \gamma \rightarrow J/\psi  X$, but rather the 
color-singlet mechanism. And it was also found in ref.~\cite{Qiao:2003ba} that 
the QCD higher order process $\gamma \gamma \rightarrow J/\psi  c \bar{c}$ 
gives the same order and even larger contribution at high $p_t$ than  
the leading order color singlet processes. 
In ref.~\cite{Hagiwara:2007bq}, at NLO the process $c  g \rightarrow J/\psi  c$ where the initial $c$ quark is the intrinsic c quark from proton at Tevatron, gives larger contribution at high $p_t$ than the leading order color singlet processes. 
The large discrepancies found in the single and double charmonium production in 
$e^+e^-$ annihilation at B factories between LO theoretical predictions 
\cite{Braaten:2002fi, Liu:2002wq} and experimental results
~\cite{Abe:2002rb,Aubert:2005tj} were studied in many work.
It seems that it may be resolved by including higher order correction: NLO QCD 
and relativistic corrections
\cite{Braaten:2002fi, Zhang:2005ch, Gong:2008ce}.

Based on NRQCD, the LO calculation predicts a sizable transverse 
polarization for $\jpsi$ at high $p_t$\cite{beneke:96yr}
while the measurement at Fermilab Tevatron\cite{Abulencia:2007us} gives almost  
unpolarized result.
Recently, NLO QCD corrections to $\jpsi$ hadronproduction 
have been calculated  by Campbell, Maltoni and Tramontano\cite{Campbell:2007ws}.
The results show that the total cross section is boosted by a factor of 
about 2 and the $\jpsi$ transverse momentum $p_t$ distribution 
is enhanced more and more as $p_t$ becomes larger.
A real correction process $g g\rightarrow \jpsi c \overline{c}$ at NLO 
was calculated by Artoisenet, 
Lansberg and Maltoni\cite{Artoisenet:2007xi}. It gives sizable contribution
to $p_t$ distribution of $\jpsi$ at high $p_t$ region, and it alone gives
almost unpolarized result. A s-channel treatment to $\jpsi$ hadronproduction gives longitudinal polarization
by H. Haberzettl and J. P. Lansberg\cite{Haberzettl:2007kj}
Therefore it is very interesting to know the result of $\jpsi$ polarization 
when NLO QCD corrections are included. In this letter, we calculate the 
NLO QCD corrections to the $\jpsi$ polarization in hadronproduction at Tevatron and LHC.
In the calculation, we use our Feynman Diagram Calculation package (FDC)\cite{FDC} 
with newly added part of a complete set of method to calculate tensor and scalar integrals 
in dimensional regularization, which was used in our previous work\cite{Gong:2008ce}.

For LO process $g(p_1)+ g(p_2) \rightarrow \jpsi(p_3) + g(p_4)$, by using the NRQCD factorization 
formalism, the partonic cross section is expressed as 
\be
\dfrac{\mathrm{d}\hat{\s}^{B}}{\mathrm{d}t}=
\dfrac{5\pi\a_s^3|R_s(0)|^2[s^2(s-1)^2+t^2(t-1)^2+u^2(u-1)^2]} {144m_c^5s^2(s-1)^2(t-1)^2(u-1)^2} ,
\ee 
with
\be
s=\dfrac{(p_1+p_2)^2}{4m_c^2},\quad t=\dfrac{(p_1-p_3)^2}{4m_c^2}, \quad u=\dfrac{(p_1-p_4)^2}{4m_c^2},\NO  
\ee
where $R_s(0)$ is the radial wave function at the origin of $\jpsi$ and the 
approximation $M_{\jpsi}=2m_c$ is taken.
The LO total cross section is obtained by convoluting the partonic cross section 
with the parton distribution function (PDF) $G_g(x,\mu_f)$ in the proton:
\be
\s^B=\int \mathrm{d}x_1\mathrm{d}x_2 G_g(x_1,\mu_f)G_g(x_2,\mu_f)\hat{\s}^B ,
\ee 
where $\mu_f$ is the factorization scale. In the following, $\hat{\s}$ represents corresponding partonic cross section.

The NLO contribution to the process can be separated into the virtual corrections, 
arising from loop diagrams, and the real corrections, arising from the radiation 
of a real gluon or a light (anti)quark or the final-state gluon splits into 
light quark-antiquark pairs. There are 
UV, IR and Coulomb singularities in the calculation of the virtual
corrections. UV-divergences from self-energy and triangle diagrams are canceled 
upon the renormalization of the QCD gauge coupling constant, the charm quark 
mass and field, and the gluon field. Here we adopt same renormalization scheme 
as ref.~\cite{Klasen:2004tz}. The renormalization constant of charm quark mass $Z_m$ and field $Z_2$, and gluon field $Z_3$ are defined in the on-mass-shell(OS) 
scheme while that of QCD gauge coupling $Z_g$ is defined in the 
modified-minimal-subtraction($\overline{\mathrm{MS}}$) scheme:
\bea
\delta Z_m^{OS}&=&-3C_F\dfrac{\alpha_s}{4\pi}\left[\dfrac{1}{\e_{UV}} -\gamma_E +\ln\dfrac{4\pi \mu_r^2}{m_c^2} +\frac{4}{3}\right] ,\NO \\
\delta Z_2^{OS}&=&-C_F\dfrac{\alpha_s}{4\pi}\left[\dfrac{1}{\e_{UV}} +\dfrac{2}{\e_{IR}} -3\gamma_E +3\ln\dfrac{4\pi \mu_r^2}{m_c^2} +4 \right] ,\NO \\
\delta Z_3^{OS}&=&\dfrac{\alpha_s}{4\pi}\left[(\beta_0-2C_A)\left(\dfrac{1}{\e_{UV}} -\dfrac{1}{\e_{IR}}\right)
\right] , \\
\delta Z_g^{\overline{\mathrm{MS}}}&=&-\dfrac{\beta_0}{2}\dfrac{\alpha_s}{4\pi}\left[\dfrac{1}{\e_{UV}} -\gamma_E +\ln(4\pi)\right] \NO.
\eea
Where $\g_E$ is Euler's constant, $\b_0=\frac{11}{3}C_A-\frac{4}{3}T_Fn_f$ is the 
one-loop coefficient of the QCD beta function and $n_f$ is the number of active 
quark flavors. There are three massless light quarks $u, d, s$, so $n_f$=3. In $SU(3)_c$, color factors are given by 
$T_F=\frac{1}{2}, C_F=\frac{4}{3}, C_A=3$. And $\mu_r$ is the renormalization scale.

\begin{figure}
\center{
\includegraphics*[scale=0.33]{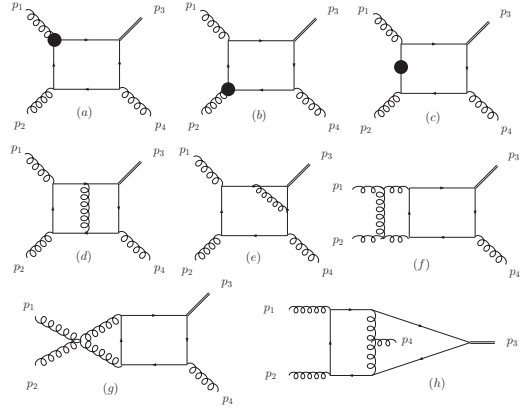}
\caption {\label{fig:NLO}One-loop diagrams for $g g \rightarrow \jpsi g$. 
Group (a) and (b) are 
counter-term diagrams of the quark-gluon vertex and corresponding loop diagrams, 
Group (c) are the quark self-energy diagrams and corresponding 
counter-term ones. More diagrams can be obtained by permutation of 
external gluons. }}
\end{figure}

After having fixed the renormalization scheme, there are 129 NLO diagrams 
,including counter-term diagrams,
which are shown in Fig.~\ref{fig:NLO}, divided into 8 groups. Diagrams of group 
$(e)$ that has a virtual gluon line connected with the quark pair lead to 
Coulomb singularity $\sim \pi^2/v$, which can be isolated by introducing 
a small relative velocity $v=|\vec{p}_{c}-\vec{p}_{\bar{c}}|$ and mapped into the $c\bar{c}$ 
wave function. 
%
By adding all diagrams together, the virtual corrections to the 
differential cross section can be connected to the virtual amplitude as
\be
\dfrac{\mathrm{d}\hat{\s}^{V}}{\mathrm{d}t} \propto 2\mathrm{Re}(M^BM^{V*}),
\ee
where $M^B$ is the amplitude at LO, and $M^V$ is the renormalized 
amplitude at NLO and is UV and Coulomb finite, 
but it still contains the IR divergences:
\bea
M^V|_{IR}=\frac{\a_s}{2\pi} \frac{\Gamma(1-\e)}{\Gamma(1-2\e)} \left(\frac{4 \pi \mu_r^2}{s_{12}}\right)^{\e}\bigl[-\dfrac{9}{2\e^2} -\dfrac{3}{2\e}\NO\\ 
\times\bigl(\ln\dfrac{s}{-t} +\ln\dfrac{s}{-u} -\dfrac{1}{3}n_f +\dfrac{11}{2}\bigr)
\bigr] M^B.
\eea

The real corrections arise from processes 
$gg\rightarrow \jpsi gg$,
$gg\rightarrow \jpsi q\overline{q}$ and 
$gq(\overline{q})\rightarrow\jpsi gq(\overline{q})$.
The phase space integration of above processes will generate IR singularities, 
which are either soft or collinear and can be conveniently isolated 
by slicing the phase space into different regions. We use the two-cutoff 
phase space slicing method \cite{Harris:2001sx}, which introduces two small 
cutoffs to decompose the phase space into three parts. And
then the real cross section could be written as
\be
\s^R=\s^{H\overline{C}}+\s^S+\s^{HC}+\s^{HC}_{add}.
\ee
The hard noncollinear part $\s^{H\overline{C}}$ is IR finite 
and is numerically computed using standard Monte-Carlo integration techniques. 
$\hat{\s}^S$ from the soft regions contains 
soft singularities and is calculated analytically under soft approximation.
It is easy to find that soft singularities for a gluon emitted from the 
charm quark pair in the S-wave color singlet $\jpsi$ are canceled 
by each other. 
$\s^{HC}$ from the hard collinear regions contains 
collinear singularities which is factorized and partly absorbed into the redefinition 
of the PDF (usually called mass factorization \cite{Altarelli:1979ub}). 
Here we adopt a scale dependent PDF using the $\overline{\rm MS}$ 
convention given by \cite{Harris:2001sx}. After the  
redefinition of the PDF, an additional term $\s^{HC}_{add}$ is separated out.
Finally, all the IR singularities are canceled analytically for 
$\hat{\s}^S+\hat{\s}^{HC}+\hat{\s}^{V}$.
And it is found that only one color factor $d_{abc}$ appears in both $M^B$ 
and $M^V$ with $a,b$ and $c$ being the color index of the three gluons in the process.

To obtain the transverse momentum distribution of $\jpsi$, a transformation for integration variables ($\md x_2 \md t \rightarrow J\md p_t \md y$) is introduced. Thus we have
\bea
\dfrac{\md \s}{\md p_t}= 
\int J \md x_1 \md y G_g(x_1,\mu_f)G_g(x_2,\mu_f) \dfrac{\md \hat \s}{\md t},
\eea
where $y$ and $p_t$ is the rapidity and transverse momentum of $\jpsi$ in laboratory frame respectively.
The polarization measurement $\alpha$ is defined as: 
\be
\alpha(p_t)=\frac{{\md\s_T}/{\md p_t}-2 {\md\s_L}/{\md p_t}}
                 {{\md\s_T}/{\md p_t}+2 {\md\s_L}/{\md p_t}}.
\ee
It represents the measurement of $\jpsi$ polarization as function 
of $p_t$. To calculate $\alpha(p_t)$, the polarization of $\jpsi$ must be kept 
in the calculation. 
The partonic differential cross section with $\jpsi$ polarized could be expressed explicitly as:
\be
\dfrac{\md \hat{\s}_{\lambda}}{\md t}= a~\epsilon(\lambda) \cdot \epsilon^*(\lambda) + \sum_{i,j=1,2} a_{ij} ~p_i \cdot \epsilon(\lambda) ~p_j \cdot \epsilon^*(\lambda),
\label{eqn:polar}
\ee
where $\lambda=T_1,T_2,L$. $\epsilon(T_1),~\epsilon(T_2),~\epsilon(L)$ are the two 
transverse polarization vectors and the longitude one for $\jpsi$,
and the polarization of all the other particles are summed up in n-dimensions. 
It causes more difficult tensor reduction path than that with all the 
polarization being summed over in virtual correction calculation. It is founded that 
$a$ and $a_{ij}$ are finite when the virtual correction and real correction are summed up. 

For $\s^{HC}_{add}$ and $\hat{\s}^S+\hat{\s}^{HC}+\hat{\s}^{V}$, the calculation is done
in Eq.~(\ref{eqn:polar}) as well as in the usual way in which 
the polarization for all particles are summed up.
The two results are used to check with each other numerically.
In the third way, the polarization of gluon is also kept and used to check gauge invariance 
by replacing the gluon polarization vector
to the gluon's 4-momentum in the final numerical calculation.
To calculate $\s^{H\bar{C}}$ in real correction processes, the numerical amplitude
calculation is used and the polarization of gluons
are only summed for their physical freedom to avoid the involving of diagrams with 
external ghosts lines.
\begin{figure}
\center{
\includegraphics*[scale=0.34]{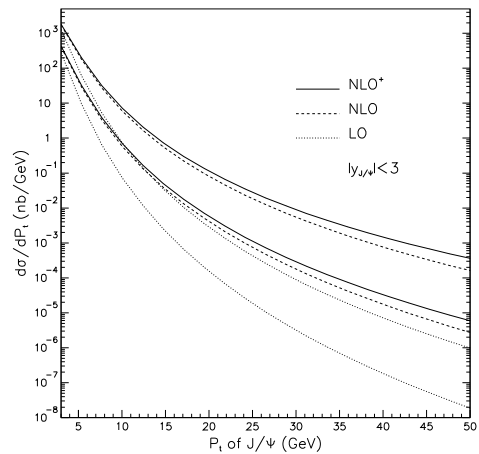}
\caption {\label{fig:pt}Transverse momentum distribution of differential 
cross section with $\mu_r=\mu_f=\sqrt{(2m_c)^2+p_t^2}$ at LHC (upper curves) 
and Tevatron (lower curves). Center mass energy are $\sqrt{s}_{\rm Tevatron}=1.98$ TeV 
and $\sqrt{s}_{\rm LHC}=14$ TeV.
$\rm NLO^+$ denotes result including contribution from
$gg\rightarrow \jpsi c\overline{c}$ at NLO.}}
\end{figure}

In our numerical calculations, the CTEQ6L1 and CTEQ6M PDFs, and the corresponding fitted value for $\a_s(M_Z)=0.130$ and $\a_s(M_Z)=0.118$, are used for LO and NLO predictions respectively. For the charm quark mass and the wave function at the 
origin of $\jpsi$, $m_c=1.5 \gev$ and $|R_s(0)|^2 =0.810 \gev^3$ are used. 
The two phase space cutoffs $\d_s$ and $\d_c$ are chosen as $\d_s=10^{-3}$ 
and $\d_c=\d_s/50$ as default choice. To check the two cutoffs invariance for the
final results, different values of $\d_s$ and $\d_c$, down to $\d_s=10^{-5}$, 
are used and the invariance is found within the error control (less than one percent).
It is known that the perturbative expansion calculation is not applicable to the regions 
with small transverse momentum and large rapidity of $\jpsi$. Therefore,
the result is restricted in the domain $p_t>3$ GeV and $|y_{\jpsi}|$ less than 3 or 0.6.

The dependence of the total cross section at the renormalization scale $\mu_r$ and 
factorization scale $\mu_f$ are obtained and it agrees with the Fig.3 in ref~\cite{Campbell:2007ws}.
In Fig.~\ref{fig:pt}, the $p_t$ distribution of $\jpsi$ is shown.  
\begin{figure*}
\center{
\includegraphics*[scale=0.35]{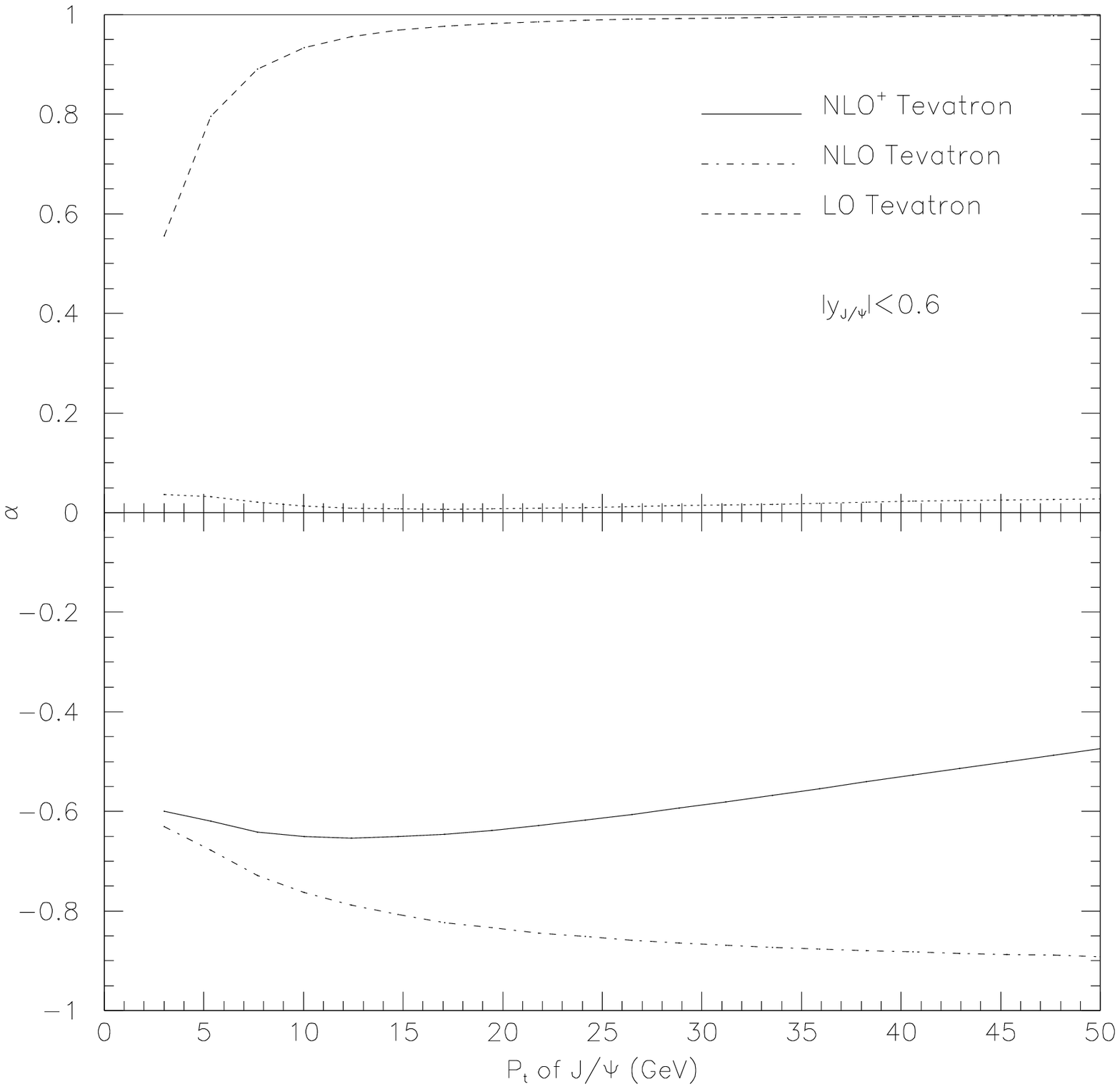}
\includegraphics*[scale=0.35]{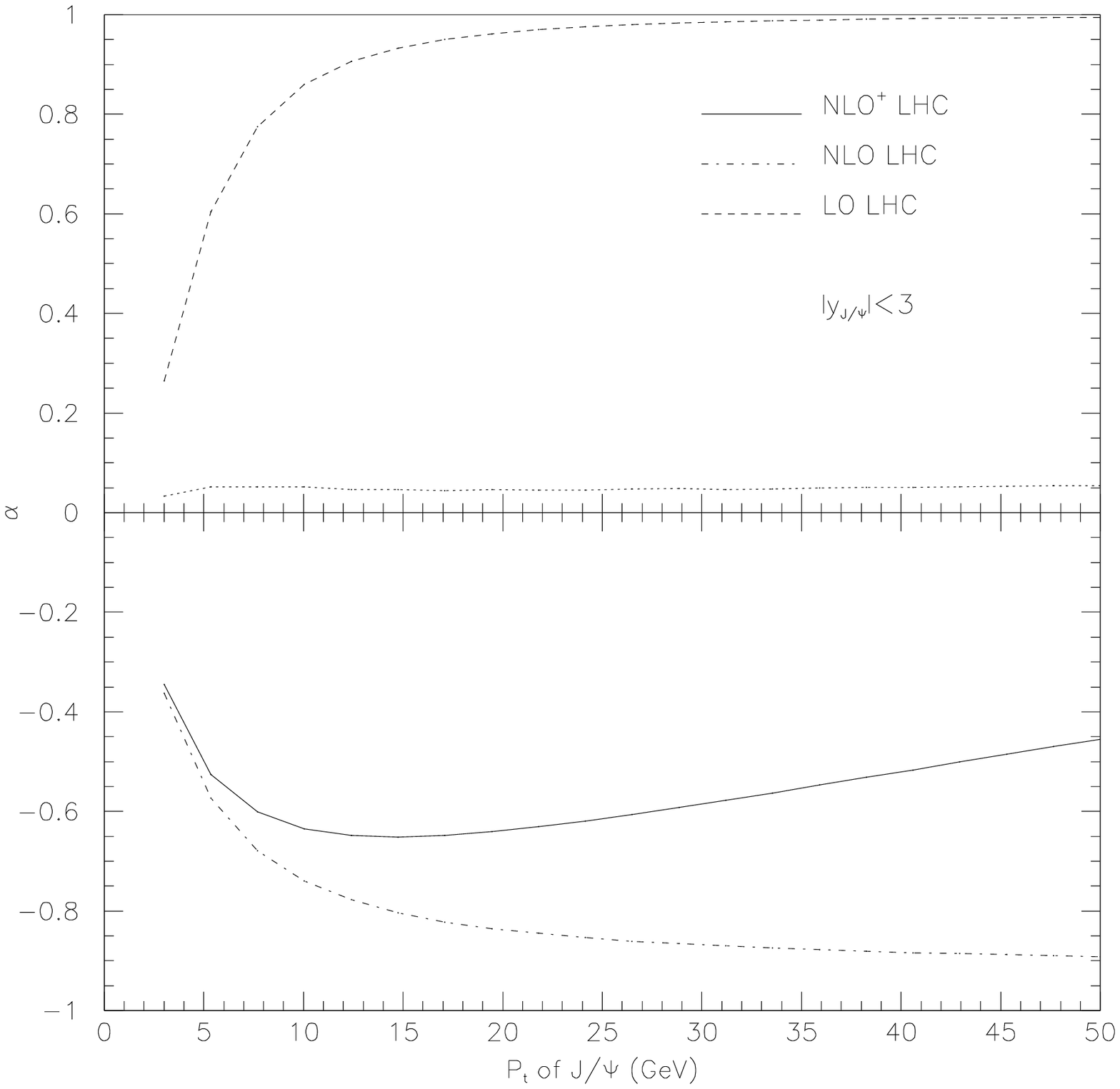}
\caption {\label{fig:polar_lhc}Transverse momentum distribution of polarization 
$\a$ with $\mu_r=\mu_f=\sqrt{(2m_c)^2+p_t^2}$ at Tevatron(left) and LHC(right).
The unlabeled dotted line denotes 
the polarization of $gg\rightarrow \jpsi c\overline{c}$ 
and $\rm NLO^+$ denotes result including that of $gg\rightarrow \jpsi c\overline{c}$.}}
\end{figure*}
The $p_t$ distribution of $\jpsi$ polarization $\a$ is shown 
in Fig.~\ref{fig:polar_lhc}. At LO, $\a$ is 
always positive and becomes closer to 1 as $p_t$ increases from 3 GeV to 50 GeV, 
which indicates that transverse polarization is always larger than longitude
polarization and transverse polarization plays 
the major role in high $p_t$ region. But there is extremely change when including 
NLO QCD corrections. $\a$ is always negative and becomes closer to -0.9 as $p_t$ 
increases from 3 GeV to 50 GeV, which indicates that transverse polarization is always 
smaller than longitude polarization and longitude polarization plays the major role
in high $p_t$ region. Meanwhile the $\jpsi$ 
polarization of process $gg\rightarrow \jpsi c\overline{c}$ is near zero. 
By including contribution of this subprocess, the result shown 
in the figures as $\rm NLO^+$ is closer to the experimental result.

In conclusion,
we calculated the NLO QCD correction of $\jpsi$ hadronproduction at Tevatron 
and LHC. The method of dimensional regularization is taken to deal with the UV 
and IR singularities in the calculation, and the Coulomb singularity is 
isolated and absorbed into the $c\bar{c}$ bound state wave function. To deal 
with the soft and collinear singularities, the two-cutoff phase space slicing 
method is used in the calculation of real corrections. After adding all 
contribution together, a result which is UV, IR and Coulomb finite is obtained. 
Numerically, we obtain a K factor of total cross section 
(ratio of NLO to LO) of about 2 at $\mu_r=\mu_f=\sqrt{(2m_c)^2+p_t^2}$. 
The transverse momentum distribution 
of $\jpsi$ is presented and it shows that the NLO corrections would 
boost the differential cross section more and more as $p_t$ becomes larger 
and reaches about 2 or 3 order of magnitude at $p_t=50 \gev$. It confirms the calculation 
in Ref.~\cite{Campbell:2007ws}. The real corrections for $gg\rightarrow \jpsi c\overline{c}$ is also calculated 
and agree with those of Ref.~\cite{Artoisenet:2007xi}.
 
The $\jpsi$ polarization at NLO is studied for the first time and the 
results show that the $\jpsi$ polarization is extremely changed from 
more transversal polarization at LO into more longitudinal polarization at NLO. 
Although it gives more longitudinal polarization than the recent experimental 
result \cite{Abulencia:2007us} on the $\jpsi$ polarization at Tevatron. 
It sheds light on the solution to the large discrepancy of $\jpsi$ polarization 
between LO theoretical predication and the experimental measurement, and suggests 
that the next important step is to calculate the NLO correction for color 
octet state $\jpsi^{(8)}$ hadronproduction.  To re-fix the color-octet 
matrix elements and to see what happens to the polarization of $\jpsi$ at NLO.  

This work is supported by the National Natural Science Foundation of 
China (No.~10775141) and by the Chinese Academy of Sciences under 
Project No. KJCX3-SYW-N2.

\bibliography{paper.bbl}

\begin{thebibliography}{30}
\expandafter\ifx\csname natexlab\endcsname\relax\def\natexlab#1{#1}\fi
\expandafter\ifx\csname bibnamefont\endcsname\relax
  \def\bibnamefont#1{#1}\fi
\expandafter\ifx\csname bibfnamefont\endcsname\relax
  \def\bibfnamefont#1{#1}\fi
\expandafter\ifx\csname citenamefont\endcsname\relax
  \def\citenamefont#1{#1}\fi
\expandafter\ifx\csname url\endcsname\relax
  \def\url#1{\texttt{#1}}\fi
\expandafter\ifx\csname urlprefix\endcsname\relax\def\urlprefix{URL }\fi
\providecommand{\bibinfo}[2]{#2}
\providecommand{\eprint}[2][]{\url{#2}}

\bibitem[{\citenamefont{Braaten and Fleming}(1995)}]{Braaten:1994vv}
\bibinfo{author}{\bibfnamefont{E.}~\bibnamefont{Braaten}} \bibnamefont{and}
  \bibinfo{author}{\bibfnamefont{S.}~\bibnamefont{Fleming}},
  \bibinfo{journal}{Phys. Rev. Lett.} \textbf{\bibinfo{volume}{74}},
  \bibinfo{pages}{3327} (\bibinfo{year}{1995}).

\bibitem[{\citenamefont{Bodwin et~al.}(1995)\citenamefont{Bodwin, Braaten, and
  Lepage}}]{Bodwin:1994jh}
\bibinfo{author}{\bibfnamefont{G.~T.} \bibnamefont{Bodwin}},
  \bibinfo{author}{\bibfnamefont{E.}~\bibnamefont{Braaten}}, \bibnamefont{and}
  \bibinfo{author}{\bibfnamefont{G.~P.} \bibnamefont{Lepage}},
  \bibinfo{journal}{Phys. Rev.} \textbf{\bibinfo{volume}{D51}},
  \bibinfo{pages}{1125} (\bibinfo{year}{1995}).

\bibitem{Ko:1996xw}
P. Ko, J. Lee and H.~S. Song, \prd{\bf 54}, 4312 (1996);
B.~A. Kniehl and G. Kramer, \plb{\bf 413}, 416 (1997).

\bibitem{kramer:1995nb}
M. Kramer \npb{\bf 459}, 3 (1996);
M. Kramer, J. Zunft, J. Steegborn and P.~M. Zerwas, \plb{\bf 348}, 657 (1995);
M. Cacciari and M. Kramer, \prl{\bf 76}, 4128 (1996);
J. Amundson, S. Fleming and I. Maksymyk, \prd{\bf 56}, 5844 (1997).

\bibitem[{\citenamefont{Klasen et~al.}(2002)\citenamefont{Klasen, Kniehl,
  Mihaila, and Steinhauser}}]{Klasen:2001cu}
\bibinfo{author}{\bibfnamefont{M.}~\bibnamefont{Klasen}},
  \bibinfo{author}{\bibfnamefont{B.~A.} \bibnamefont{Kniehl}},
  \bibinfo{author}{\bibfnamefont{L.~N.} \bibnamefont{Mihaila}},
  \bibnamefont{and}
  \bibinfo{author}{\bibfnamefont{M.}~\bibnamefont{Steinhauser}},
  \bibinfo{journal}{Phys. Rev. Lett.} \textbf{\bibinfo{volume}{89}},
  \bibinfo{pages}{032001} (\bibinfo{year}{2002}).

\bibitem[{\citenamefont{de~Boer and Sander}(2004)}]{deBoer:2003xm}
\bibinfo{author}{\bibfnamefont{W.}~\bibnamefont{de~Boer}} \bibnamefont{and}
  \bibinfo{author}{\bibfnamefont{C.}~\bibnamefont{Sander}},
  \bibinfo{journal}{Phys. Lett.} \textbf{\bibinfo{volume}{B585}},
  \bibinfo{pages}{276} (\bibinfo{year}{2004}).

\bibitem{Qiao:2003ba}
C. F. Qiao and J. X. Wang, \prd{\bf 69}, 014015 (2004).

\bibitem[{\citenamefont{Hagiwara et~al.}(2007)\citenamefont{Hagiwara, Qi, Qiao,
  and Wang}}]{Hagiwara:2007bq}
\bibinfo{author}{\bibfnamefont{K.}~\bibnamefont{Hagiwara}},
  \bibinfo{author}{\bibfnamefont{W.}~\bibnamefont{Qi}},
  \bibinfo{author}{\bibfnamefont{C.~F.} \bibnamefont{Qiao}}, \bibnamefont{and}
  \bibinfo{author}{\bibfnamefont{J.~X.} \bibnamefont{Wang}}
  (\bibinfo{year}{2007}), \eprint{arXiv:0705.0803 [hep-ph]}.

\bibitem[{\citenamefont{Braaten and Lee}(2003)}]{Braaten:2002fi}
\bibinfo{author}{\bibfnamefont{E.}~\bibnamefont{Braaten}} \bibnamefont{and}
  \bibinfo{author}{\bibfnamefont{J.}~\bibnamefont{Lee}},
  \bibinfo{journal}{Phys. Rev.} \textbf{\bibinfo{volume}{D67}},
  \bibinfo{pages}{054007} (\bibinfo{year}{2003}).

\bibitem{Liu:2002wq}
K.-Y. Liu, Z.-G. He and K.-T. Chao, \plb{\bf 557}, 45 (2003);
K. Hagiwara, E. Kou and C.-F. Qiao, \plb{\bf 570}, 39 (2003).

\bibitem[{\citenamefont{Abe et~al.}(2002)}]{Abe:2002rb}
\bibinfo{author}{\bibfnamefont{K.}~\bibnamefont{Abe}} \bibnamefont{et~al.}
  (\bibinfo{collaboration}{Belle}), \bibinfo{journal}{Phys. Rev. Lett.}
  \textbf{\bibinfo{volume}{89}}, \bibinfo{pages}{142001}
  (\bibinfo{year}{2002}).

\bibitem[{\citenamefont{Aubert et~al.}(2005)}]{Aubert:2005tj}
\bibinfo{author}{\bibfnamefont{B.}~\bibnamefont{Aubert}} \bibnamefont{et~al.}
  (\bibinfo{collaboration}{BABAR}), \bibinfo{journal}{Phys. Rev.}
  \textbf{\bibinfo{volume}{D72}}, \bibinfo{pages}{031101}
  (\bibinfo{year}{2005}).

\bibitem{Zhang:2005ch}
Y.-J. Zhang, Y.-j. Gao and K.-T. Chao, \prl{\bf 96}, 092001 (2006).
Z.-G. He, Y. Fan and K.-T. Chao, \prd{\bf 75}, 074011 (2007).
Y.-J. Zhang and K.-T. Chao, \prl{\bf 98}, 092003 (2007).

\bibitem{Gong:2008ce}
B. Gong and J.-X. Wang, arXiv:0712.4220 [hep-ph], (2007);
  arXiv:0801.0648 [hep-ph], (2008)

\bibitem{beneke:96yr}
M. Beneke and I.Z. Rothstein, \plb{\bf 372}, 157 (1996),
[Erratum-ibid. B{\bf 389}, 769 (1996)];
M. Beneke and M. Kr\"amer, \prd{\bf 55}, 5269 (1997).
E. Braaten, B.A. Kniehl, and J. Lee,
\prd{\bf 62}, 094005 (2000);
B. A. Kniehl and J. Lee, \prd{\bf 62}, 114027 (2000).
A.~K. Leibovich, \prd{bf 56}, 4412 (1997).

\bibitem[{\citenamefont{Abulencia et~al.}(2007)}]{Abulencia:2007us}
\bibinfo{author}{\bibfnamefont{A.}~\bibnamefont{Abulencia}}
  \bibnamefont{et~al.} (\bibinfo{collaboration}{CDF}), \bibinfo{journal}{Phys.
  Rev. Lett.} \textbf{\bibinfo{volume}{99}}, \bibinfo{pages}{132001}
  (\bibinfo{year}{2007}).

\bibitem[{\citenamefont{Campbell et~al.}(2007)\citenamefont{Campbell, Maltoni,
  and Tramontano}}]{Campbell:2007ws}
\bibinfo{author}{\bibfnamefont{J.}~\bibnamefont{Campbell}},
  \bibinfo{author}{\bibfnamefont{F.}~\bibnamefont{Maltoni}}, \bibnamefont{and}
  \bibinfo{author}{\bibfnamefont{F.}~\bibnamefont{Tramontano}},
  \bibinfo{journal}{Phys. Rev. Lett.} \textbf{\bibinfo{volume}{98}},
  \bibinfo{pages}{252002} (\bibinfo{year}{2007}).

\bibitem[{\citenamefont{Artoisenet et~al.}(2007)\citenamefont{Artoisenet,
  Lansberg, and Maltoni}}]{Artoisenet:2007xi}
\bibinfo{author}{\bibfnamefont{P.}~\bibnamefont{Artoisenet}},
  \bibinfo{author}{\bibfnamefont{J.~P.} \bibnamefont{Lansberg}},
  \bibnamefont{and} \bibinfo{author}{\bibfnamefont{F.}~\bibnamefont{Maltoni}},
  \bibinfo{journal}{Phys. Lett.} \textbf{\bibinfo{volume}{B653}},
  \bibinfo{pages}{60} (\bibinfo{year}{2007}).

\bibitem{Haberzettl:2007kj}
H. Haberzettl and J. P. Lansberg,
 Phys.\ Rev.\ Lett.\  {\bf 100}, 032006 (2008)

\bibitem[{\citenamefont{Wang}(2004)}]{FDC}
\bibinfo{author}{\bibfnamefont{J.-X.} \bibnamefont{Wang}},
  \bibinfo{journal}{Nucl. Instrum. Meth.} \textbf{\bibinfo{volume}{A534}},
  \bibinfo{pages}{241} (\bibinfo{year}{2004}).

\bibitem[{\citenamefont{Klasen et~al.}(2005)\citenamefont{Klasen, Kniehl,
  Mihaila, and Steinhauser}}]{Klasen:2004tz}
\bibinfo{author}{\bibfnamefont{M.}~\bibnamefont{Klasen}},
  \bibinfo{author}{\bibfnamefont{B.~A.} \bibnamefont{Kniehl}},
  \bibinfo{author}{\bibfnamefont{L.~N.} \bibnamefont{Mihaila}},
  \bibnamefont{and}
  \bibinfo{author}{\bibfnamefont{M.}~\bibnamefont{Steinhauser}},
  \bibinfo{journal}{Nucl. Phys.} \textbf{\bibinfo{volume}{B713}},
  \bibinfo{pages}{487} (\bibinfo{year}{2005}).

\bibitem[{\citenamefont{Harris and Owens}(2002)}]{Harris:2001sx}
\bibinfo{author}{\bibfnamefont{B.~W.} \bibnamefont{Harris}} \bibnamefont{and}
  \bibinfo{author}{\bibfnamefont{J.~F.} \bibnamefont{Owens}},
  \bibinfo{journal}{Phys. Rev.} \textbf{\bibinfo{volume}{D65}},
  \bibinfo{pages}{094032} (\bibinfo{year}{2002}).

\bibitem[{\citenamefont{Altarelli et~al.}(1979)\citenamefont{Altarelli, Ellis,
  and Martinelli}}]{Altarelli:1979ub}
\bibinfo{author}{\bibfnamefont{G.}~\bibnamefont{Altarelli}},
  \bibinfo{author}{\bibfnamefont{R.~K.} \bibnamefont{Ellis}}, \bibnamefont{and}
  \bibinfo{author}{\bibfnamefont{G.}~\bibnamefont{Martinelli}},
  \bibinfo{journal}{Nucl. Phys.} \textbf{\bibinfo{volume}{B157}},
  \bibinfo{pages}{461} (\bibinfo{year}{1979}).

\end{thebibliography}
\end{document}